# Covid-19 infodemic reveals new tipping point epidemiology and a revised *R* formula

N.F. Johnson[1,2], N. Velásquez[2], O.K. Jha[2], H. Niyazi[2], R. Leahy[2,3], N. Johnson Restrepo[2,3], R. Sear[4], P. Manrique[5], Y. Lupu[6], P. Devkota[7], S. Wuchty[7]
[1]*Physics Department, George Washington University, Washington D.C. 20052*
[2]*Institute for Data, Democracy and Politics, George Washington University, Washington D.C. 20052*
[3]*ClustrX LLC, Washington D.C.*
[4]*Department of Computer Science, George Washington University, Washington D.C. 20052*
[5]*Theoretical Biology and Biophysics Group, Los Alamos National Laboratory, Los Alamos, New Mexico, 87545*
[6]*Department of Political Science, George Washington University, Washington D.C. 20052*
[7]*Department of Computer Science, University of Miami, Miami, Florida 33146*

**Many governments have managed to control their COVID-19 outbreak with a simple message: keep the effective '*R* number' *R*<1 to prevent widespread contagion and flatten the curve. This raises the question whether a similar policy could control dangerous online 'infodemics' of information, misinformation and disinformation[1,2,3,4,5,6,7,8,9,10,11,12,13]. Here we show, using multi-platform data from the COVID-19 infodemic, that its online spreading instead encompasses a different dynamical regime where communities and users within and across independent platforms, sporadically form temporary active links on similar timescales to the viral spreading. This allows material that might have died out, to evolve and even mutate[14,15,16]. This has enabled niche networks that were already successfully spreading hate and anti-vaccination material, to rapidly become global super-spreaders of narratives featuring fake COVID-19 treatments, anti-Asian sentiment and conspiracy theories. We derive new tools that incorporate these coupled social-viral dynamics, including an online *R*, to help prevent infodemic spreading at all scales: from spreading across platforms (e.g. Facebook, 4Chan) to spreading within a given subpopulation, or community, or topic. By accounting for similar social and viral timescales, the same mathematical theory also offers a quantitative description of other unconventional infection profiles such as rumors spreading in financial markets and colds spreading in schools.**

Despite the many physical differences between a piece of information on social media and a biological virus, the term 'infodemic' adopted by the World Health Organization is attractive since it suggests importing policies from epidemiology[1,17,18,19,20] Starbird et al.'s continually updated literature review[3] spells out the dangers of infodemics of online information, misinformation, and disinformation[21,22,23,24,25,26]. The current infodemic now extends beyond COVID-19 topics such as lockdowns, protests and masks to include the 2020 U.S. elections. The cross-platform data that we have compiled of such COVID-related material from Dec 2019-June 2020, shows that the online activity ('infection') at all scales is typically very different to the single-peak infection profile and diffusive spreading predicted by standard epidemiological models (Figs. 1,2, Extended Data Figs. 3 and 5). Hence, there is a need to develop a quantitative science of such 'infodemiology' to guide any adaptation of policies from epidemiology.

Figure 1a illustrates a key online mechanism that is essentially absent from offline epidemics: online communities such as Facebook Pages or VKontakte Groups, each of which is a cluster (i.e. node) with up to a million or more members, frequently form links to other online communities both across and within platforms on a similar timescale to the effective infection and recovery times for viral material, e.g. a few days. When combined with occasional link-breaking, either intentional or by moderators, these link dynamics generate a perpetual reclustering-of-clusters within and across platforms and hence across languages and continents[27]. This means that although interest in a particular piece of (mis/dis)information may be waning within a given community, it can get injected into new communities where it can revive, evolve and mutate. Crucially, no continuous path need exist at any given time across the entire social media multiverse. Instead, paths that are piecewise in time can open up from sporadic sequences of links. Extended Data Figure 1 gives an explicit example. This is akin to crossing a wide river by serendipitous positioning of a short plank to bridge adjacent rocks. An immediate consequence is that subnetworks that were already adept at spreading contentious material under these conditions prior to 2020, including those for health misinformation (Fig. 2a) and hate (Fig. 1a), acted as early global super-spreaders for COVID-19 narratives (Fig. 1a inset), including anti-Asian racism and dis/misinformation about COVID-19 causes, cures and vaccines.

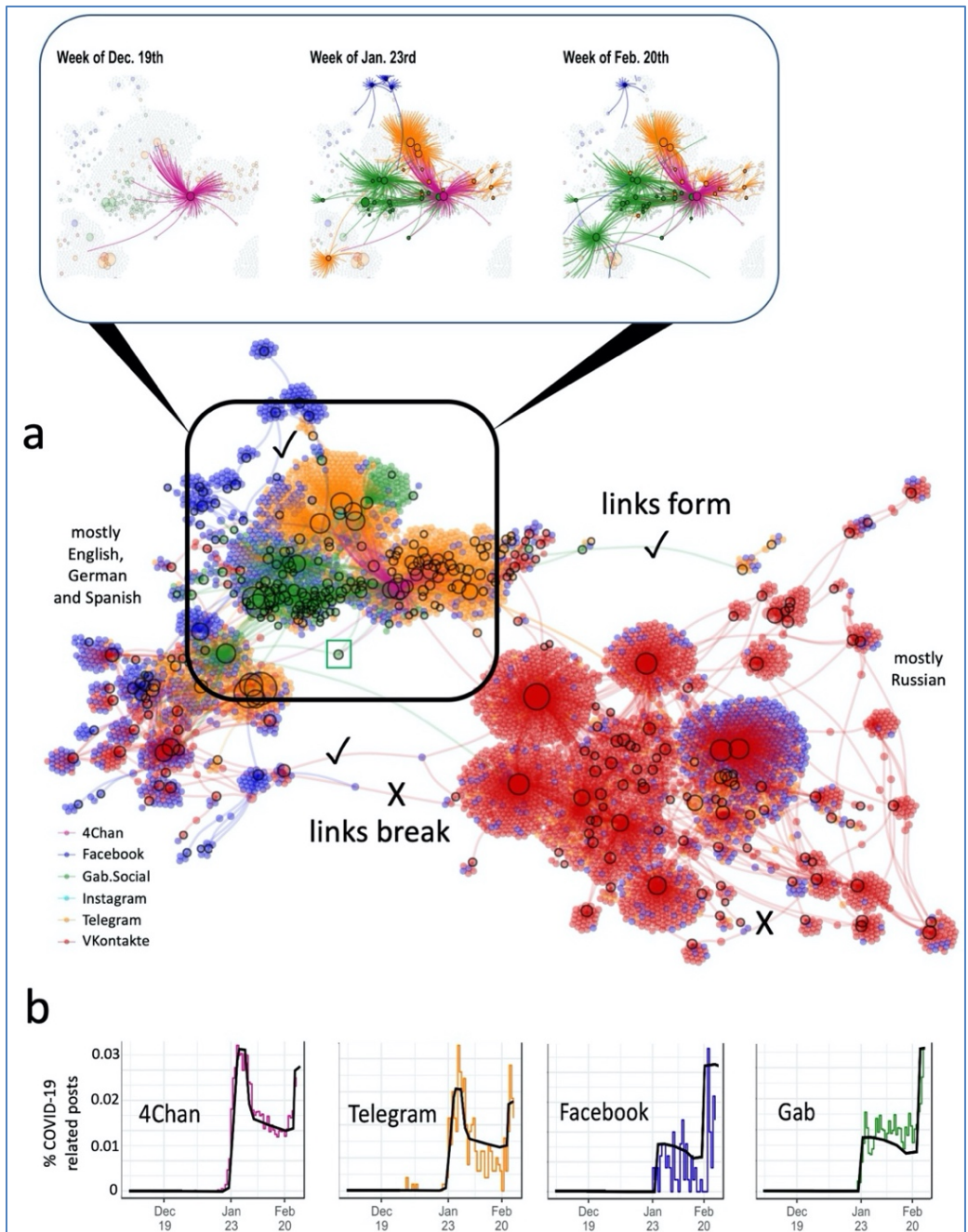

**Figure 1: Transient pathways within and across platforms for spreading. a:** Each node is a cluster of members (e.g. Gab community). For simplicity, we only show clusters whose narratives focus around hate (black circles) and clusters they link into (no rings), since these became a strong conduit for COVID-19 misinformation from December 2019 onwards (inset, top). Snapshot is from the start of COVID-19 infodemic (Jan 2020): it involves ~10,000,000 users across languages and continents who formed themselves into ~6,000 inter-linked clusters. Small green square shows example of a community that suddenly emerged and trafficked COVID-19 misinformation (see Methods). **b:** Infection profiles of COVID-19 activity feature multiple peaks, unlike standard epidemiology. Values are non-zero but too small to see on this scale prior to the major infodemic onset in mid-January. Solid black curves show the theory's predictions (SI) without employing any parameter optimization to improve fitting. Left two theory curves are identical, just rescaled in magnitude, and so are the right two theory curves: the only difference in parameter values is left two theory curves have a recovery time per cluster which is 0.5 times that for the right two theory curves. These profiles illustrate how time-dependent links allow the 'infection' to die down but then re-emerge.

The Supplementary Information (SI) details our simple approach to incorporating these missing social-viral dynamics into standard epidemiology, starting from fundamental equations. The resulting theory's predictions are in good agreement with full-scale computer simulations (Extended Data Fig. 2) and produce infection profiles consistent with those observed across all online scales (e.g. Fig. 1b and Extended Data Fig. 3). Solving these equations yields a new online 'R number' $R_{online}$ and a tipping point condition to prevent spreading across $m$ online universes (e.g. $m$ social media platforms):

$$R_{online} = \left[\frac{\sum_{i,j=1}^{m} n_i n_j f_{ij}}{\sum_{i=1}^{m} n_i b_i}\right] \frac{q_i}{q_r} < 1 \qquad (1)$$

where all the quantities on the right side of Eq. 1 are physical quantities that can be counted directly from the data, e.g. Fig. 1a. $n_i$ and $n_j$ are the fraction of clusters (e.g. Facebook Pages) in universes (platforms) $i$ and $j$ respectively. In Fig. 1a, for example, $n_i$ for $i$=VKontakte is the fraction of red nodes. $f_{ij}^{-1}$ is the typical (technically, the average) time it takes for a link to form between a cluster in $i$ and a cluster in $j$, e.g. if $i$ and $j$ are both on Facebook (FB), a link is created when FB Page A (i.e. cluster A) 'likes' FB Page B (i.e. cluster B) and hence content is more likely to be shared, though Eq. 1 also applies to other link definitions. Our data shows this is on the scale of days. $b_i^{-1}$ is the typical time it takes for the links of a cluster in universe $i$ to be removed. $C = q_i/q_r$ is the contact ratio and it is independent of the cluster network: $q_i^{-1}$ is the typical time it takes for a cluster to pass a piece of material to another cluster to which it is already linked, which our data shows is also on the scale of days, while $q_r^{-1}$ is the typical time it takes until new content added to a cluster does not mention that material. In the mass-action limit, $R_{online} \to [N] q_i/q_r$ giving the familiar $R$ number for a real virus in a contact population size $N$, which is equivalent to the average number of people infected by each infected person. Since Eq. 1 is derived mathematically, not by empirical fit or statistical scaling, it can be generalized in a systematic way to address different what-if scenarios.

Equation 1 has a variety of consequences for potential policy-making:

First, Eq. 1 takes on a transparent form when the policy focus is on policing pairs of subpopulations, e.g. a pair of platforms such as Facebook (FB) and a platform (A) that FB may not control. If both platforms have similar gross features, $f_{FB\,FB} \sim f_{AA} \equiv f$ and $f_{FB\,A} \sim f_{A\,FB} \equiv Ff$, hence Eq. 1 predicts that Facebook could prevent spreading by ensuring the fraction of inter-platform links $F < F_{crit}$, where

$$F_{crit} = \left[\frac{b_{FB} n_{FB} + b_A n_A}{2 f n_{FB} n_A}\right] \frac{1}{C} + \left[1 - \frac{1}{2 n_{FB} n_A}\right] . \qquad (2)$$

$F_{crit}$ vs. $1/C$ is a convenient straight-line graph and $1/C$ is independent of the network. Equation 2 means that Facebook, for example, can increase $b_{FB}$ to raise $F_{crit}$ above $F$, or directly reduce $F$ by reducing inter-platform links. Hence it avoids the need for blanket 'social distancing' between Facebook and platform A, or 'quarantining' of platform A, and hence avoids claims of over-policing.

Second, Eq. 1 applies at other scales simply by changing the specification of 'cluster' and 'universe', e.g. to prevent spreading between a subpopulation on Telegram that share adult or borderline extremist content, and a subpopulation of vulnerable individuals on Facebook (e.g. minors); or to prevent spreading of COVID narratives within a platform (e.g. Facebook) from anti-vaccination communities to mainstream communities (e.g. pet-lovers)[33]. It also applies to spreading within a single community by classifying nodes as individual users who then form a cluster (i.e. single community such as a Facebook Page). Since they tend to join and leave individually, the mathematics reduces primarily to monomer dynamics. Similarly anomalous infection profiles emerge (see Extended Data Fig. 3 and SI for details).

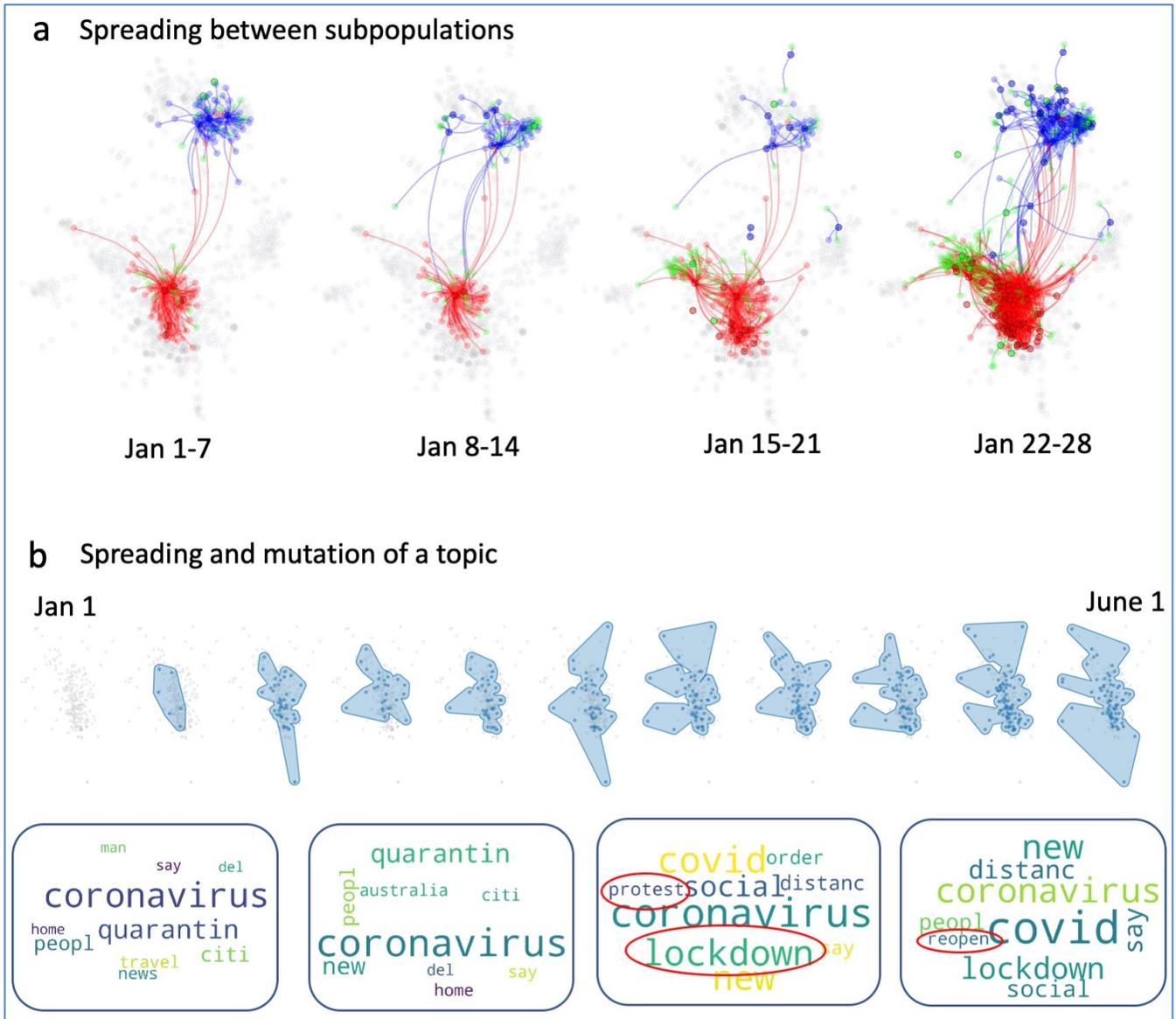

**Figure 2: Spreading within platform and topic. a:** Illustration of spreading of COVID-19 information, misinformation and disinformation between anti-vaccination (red), pro-vaccination (blue) and 'undecided' (green) clusters (each node is a Facebook Page). Each subpopulation shows similarly anomalous infection profiles to Fig. 1b and are unlike standard epidemiological diffusive spreading, e.g. the high initial spread within pro-vaccination subpopulation then decays before re-emerging. **b:** Spreading of a single topic on same network as panel a, shown at intervals of 2 weeks, is also unlike standard epidemiological diffusive spreading. This topic was identified using dynamic LDA machine-learning[34] as having the fastest rising coherence score (see SI). Word clouds illustrate how this topic's time-dependent word weightings evolve and the topic itself mutates toward protests, lockdowns and then reopenings (red rings).

Third, Eq. 1 predicts the online *herd immunity* required to prevent spreading for policy schemes that propose to 'innoculate' online[35]. The minimum fraction of clusters that if innoculated will prevent an infodemic, is estimated to be $[1 - R_{online}^{-1/3}]$. Hence $R_{online}$=10 means at least 54% of clusters require innoculation, which is similar to the value we obtain using computer simulations.

Fourth, Eq. 1 says that even if $q_i/q_r > 1$ and there is no such online 'vaccine', the prefactor can be adjusted by changing the link forming and breaking timescales, to force $R_{online} < 1$ and hence prevent the infodemic. Even if $q_i/q_r$ is not accurately known, reducing the prefactor will still lower the risk.

Fifth, Eq. 1 allows for different types of material to have different infectiousness, e.g. COVID-19 misinformation vs. racism, by having different contact ratio ($C = q_i/q_r$) values. The same is true regarding its level of truthfulness, e.g. fake news may have a higher $C$. Given the same network dynamics, an infodemic can therefore occur in one case but not another.

Sixth, the physical variables in the prefactor in Eq. 1 do not need to be the same before and after the viral material is introduced, so Eq. 1 allows for the system as a whole to subsequently react or adapt. Likewise by allowing a dependence on the contact ratio ($q_i/q_r$) or time, Eq. 1 can account for material that is so infectious that it rewires the system as it spreads, or it mutates as it spreads as in Fig. 2b.

Seventh, Eq. 1 applies for other definitions of links between clusters. The links forming or breaking between a pair of clusters may be based on the time-dependent correlations, or the amount of material shared, or the topic or nature of its content (e.g. videos vs. text) or common membership.

Eighth, since the underlying mathematical analysis centers around the derivation of the probability of links existing at a specific time, generalizations of Eq. 1 for other choices of infection process beyond SIR (Susceptible-Infected-Recovered) follow very similar mathematics.

Ninth, Eq. 1 accounts for the empirical observation that while viral material can *appear* isolated and largely eradicated on a given platform, it may have moved though inter-cluster links to other platforms where it revives before later re-emerging on the original platform. Moderators reviewing blue clusters in Fig. 1a might conclude that they have largely rid their platform of certain unwanted material in certain clusters, only to see it re-emerge at a later date in completely different clusters.

Tenth, this mathematical generalization of standard epidemiology also produces profiles similar to other real-world systems that have an interplay between social and viral timescales, e.g. online rumors in financial markets and contagious diseases (e.g. colds) within schools (see Extended Data Fig. 4).

There are limitations to our study. The 'infection' terminology will always be an imperfect analogy. While better fits to the profile shapes can be obtained by parameterizing the variables in Eq. 1, and setting-specific effects may contribute to particular profile anomalies, our focus is instead on providing a benchmark scientific framework using a minimal, parsimonious description[36]. Other social dynamics that are routine, e.g. nightly sleep, should average out globally. We have not followed specific pieces of information, e.g. the Plandemic movie, because our focus is on the system level. Our mathematical formulae are approximations, however they agree with full-scale computer simulations and reproduce key features of the empirical profiles. There will be errors in estimating the physical variables in Eq. 1, however these can be combined to give a useful error range for $R_{online}$. People may be members of multiple clusters, however our prior work found that only a small percentage are actually active across multiple clusters. Some people may avoid certain material while for others it may incite them[37], however these two types could be considered crudely as two subpopulations in Eq. 1 and 2. Clandestine agencies may contribute to spreading: however the decentralized nature of Fig. 1a makes any central control unlikely. For example, for material with a hate component, we only found a small portion of clusters linking to Kremlin-affiliated domains and these links accounted for <0.5% of all posts shared. This is consistent with claims that in-built communities (e.g. Facebook Pages) self-police for trolls and bots, as opposed to Twitter where there is no such in-built community structure.

**Methods**

The SI provides the derivation and analysis of the mathematical equations and our methods for obtaining the empirical dynamical networks of clusters and classifying their content (e.g. Fig. 1a, 2a) which follow our prior works[38,39,40,41]. Each cluster (i.e. node) in Fig. 1a is an interest-based online community (e.g. Facebook Page or VKontakte Group) and each link is a hyperlink that appears between them. We do not consider Twitter since it does not have the in-built cluster tools of Facebook etc.: such clusters are known to facilitate coordination and play a greater role in nurturing. Figure 1a includes the mainstream platforms Facebook, VKontakte, and Instagram, that have and enforce content policies, and fringe platforms with minimal content policies: Gab, Telegram, and 4Chan. Figure 1b shows typical output from our model with the same values $f_{ij} = 0.95$, $b_i = 0.05$, $q_i = 0.05$ for all four panels. For 4Chan and Telegram $q_r = 0.01$; for Gab and Facebook $q_r = 0.005$. All four fits are with these same two model profile outputs suitably scaled. Output is smoothed over timepoints like the empirical data which is collected daily. All but one node in Fig. 1a is plotted using the ForceAtlas2 algorithm, which simulates a physical system where nodes (clusters) repel each other while links act as springs, and nodes that are connected through a link attract each other. Hence nodes (clusters) closer to each other have more highly interconnected local environments while those farther apart do not. The exception is Gab group 407* ("Chinese Coronavirus", https://gab.com/groups/407*, small green square in Fig. 1a) which was manually placed to facilitate visibility. It was created in early 2020 with a focus on discussing the COVID19 pandemic: however, it immediately mixed COVID-19 fake news and hate, as well as conspiratorial content. In Fig. 1b, the vertical scale is only shown on one plot for clarity, but the others have somewhat similar scales (see Extended Data Fig. 5). Our dynamic LDA machine learning process and analysis (Fig. 2b) is described in the SI.

**Data Availability:** Dataset provided with the Supplementary Information (SI). All computer programs are given in the SI or are described fully in previous publications (see references).
**Author Contributions:** All authors contributed to the research design, the analysis, and writing the paper.
**Materials and Correspondence** should be sent to NFJ neiljohnson@gwu.edu